# Grouping Similar Seismocardiographic Signals Using Respiratory Information

Amirtaha Taebi, *Student Member*, *IEEE*, and Hansen A. Mansy
Biomedical Acoustics Research Laboratory, University of Central Florida, Orlando, FL 32816, USA
{taebi@knights., hansen.mansy@} ucf.edu

*Abstract*—Seismocardiography (SCG) offers a potential non-invasive method for cardiac monitoring. Quantification of the effects of different physiological conditions on SCG can lead to enhanced understanding of SCG genesis, and may explain how some cardiac pathologies may affect SCG morphology. In this study, the effect of the respiration on the SCG signal morphology is investigated. SCG, ECG, and respiratory flow rate signals were measured simultaneously in 7 healthy subjects. Results showed that SCG events tended to have two slightly different morphologies. The respiratory flow rate and lung volume information were used to group the SCG events into inspiratory/expiratory groups or low/high lung-volume groups, respectively. Although respiratory flow information could separate similar SCG events into two different groups, the lung volume information provided better grouping of similar SCGs. This suggests that variations in SCG morphology may be due, at least in part, to changes in the intrathoracic pressure or heart location since those parameters correlates more with lung volume than respiratory flow. Categorizing SCG events into different groups containing similar events allows more accurate estimation of SCG features, and better signal characterization, and classification.

## I. INTRODUCTION

Cardiovascular disease is a major cause of mortality in the United States as it accounts for 24.2% of total deaths [1]. Developing new technologies for cardiac monitoring and diagnosis can help improve patient management and reduce mortality. Hence, analysis of blood flow dynamics [2], [3] and the heart related signals [4] has become an active area of research. Seismocardiographic signals (SCG) are the mechanical vibrations measured noninvasively at the chest surface [5], [6]. SCG are believed to be caused by the mechanical processes associated with the heart activity (such as cardiac muscle contraction, blood momentum changes, valve closure, etc.) [7]–[9]. SCG signals contain information relating to both cardiovascular and respiratory systems [10] that might be complementary to other heart monitoring methods such as electrocardiography and phonocardiography. The SCG signals mainly contain low-frequency waves where the human auditory sensitivity is low and cannot sufficiently extract the signal characteristics accurately [11], [12]. Hence measurement and analysis of these signals may be done using computerized data acquisition and analysis, which would provide enhanced qualitative and quantitative description of the signal characteristics in both time and frequency domains [13]–[15].

SCG was previously used to estimate the respiration rate, which was found comparable to that derived from a reference respiration belt [10]. SCG signal morphology was reported to vary with different factors, including respiration cycle (inspiration vs expiration), sensor location on the chest, etc. [16], [17]. The effect of respiratory cycle has been studied [10] on some of the SCG features such as timing interval changes. However, the effect of respiration on the SCG signal morphology needs more attention [18]. During inspiration, the diaphragm moves downward, the chest wall expands, the intrathoracic pressure decreases, the lungs inflate [19], and the heart positon is displaced almost linearly with the diaphragm [20]. The decreased intrathoracic pressure increases the pulmonary blood volume, leading to an increase and decrease in the right and left atrial filling, and reduction in the left ventricular stroke volume [21]. These hemodynamic changes can affect the SCG signal morphology.

As described, the SCG variation during a respiration cycle has been mentioned before. This study, however, aims at investigating the possible physiological correlates of this morphological variation. For this purpose, the SCG events in a recording were first grouped based on the criteria that are physiologically measureable (e.g. inhalation and exhalation), and then the best criterion that could group similar SCG events together was identified. That criterion was then studied to explain why SCG morphology varies during the respiration cycle. Achievements of this study include quantification of the differences in SCG signals due to respiration, and determination of optimal respiration criterion for grouping the different SCG waveforms. Materials and methods are given in section II. Results are presented and discussed in sections III and IV, respectively, followed by conclusions in section V.

## II. METHODOLOGY

### A. Participants

The study protocol was approved by the institutional review board of the University of Central Florida, Orlando, FL. A total of 7 young individuals with no history of cardiovascular disease participated in the study after informed consent. Mean age, height, weight, body mass index (BMI), and heartbeat of the subjects were obtained and reported in Table I.

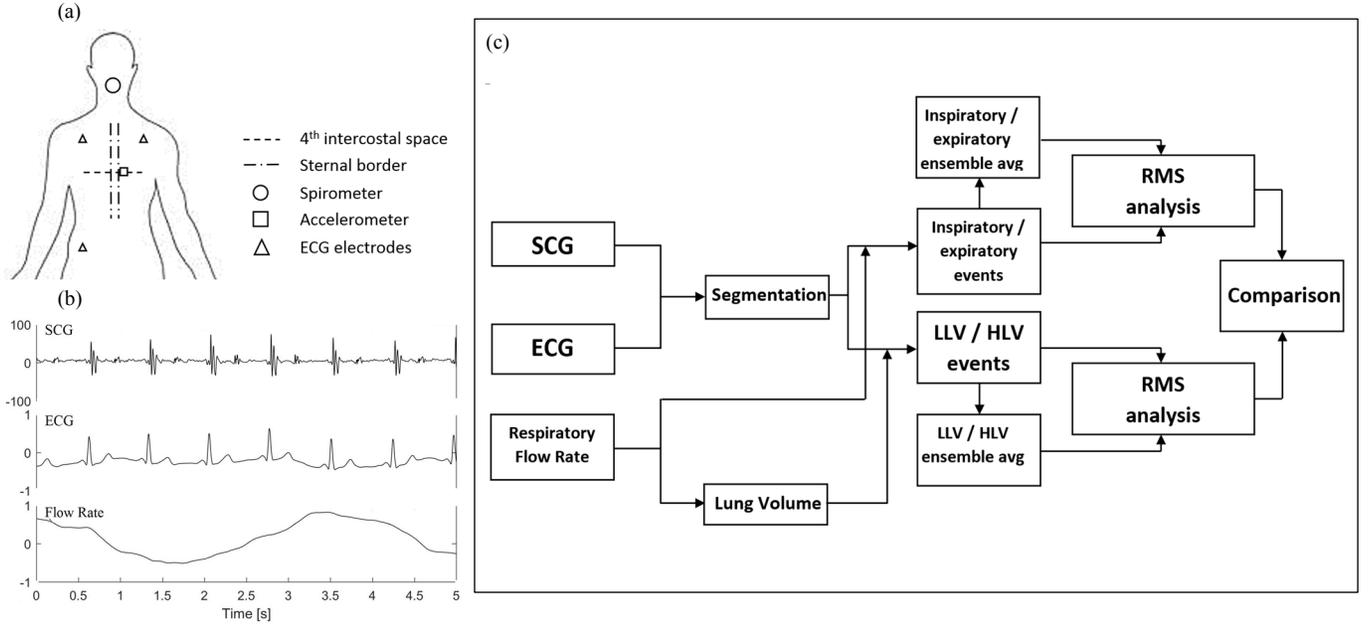

Fig. 1. (a) The location of the accelerometer, ECG electrodes and spirometer on the subject body. The accelerometer and spirometer sensors were used to measure the SCG and respiratory flow rate signals, respectively. The dashed and dash-dot lines show the 4th intercostal space and sternal border, respectively. (b) A 5 s portion of simultaneously acquired SCG, ECG, and respiratory flow rate signals. (c) Summary of the signal processing algorithm used in this study.

## B. Data Collection

All participants were instructed to lay supine on a table and breathe normally. The SCG signal was measured using a triaxial accelerometer (Model: 356A32, PCB Piezotronics, Depew, NY). The accelerometer output was amplified using a signal conditioner (Model: 482C, PCB Piezotronics, Depew, NY) with a gain factor of 100. The sensor was placed at the left sternal border and the 4th intercostal space using a double-sided tape since this location tended to have high signal-to-noise ratio. The accelerometer's x- and y-axes were aligned parallel to the anteroposterior and mediolateral directions, respectively, while the z-axis was aligned in dorso-ventral direction. In this study, the z-component of acceleration tended to be strongest, similar to previous studies [18]. Therefore, attention in the current study was focused on the analysis of this acceleration component. The respiratory flow rate of the subjects was measured using a pre-calibrated spirometer (Model: A-FH-300, iWorx Systems, Inc., Dover, NH). The flow rate signal had positive and negative amplitude during the inspiration and expiration, respectively. The voltage signal for both respiratory flow rate and SCG signals were acquired using a Control Module (Model: IX-TA-220, iWorx Systems, Inc., Dover, NH). The lung volume was calculated as the integral of the respiratory flow rate. The sensors locations are shown in Fig. 1.a.

The SCG, ECG, and respiratory signals were all measured simultaneously at a sampling frequency of 10 kHz and down-sampled to 320 Hz. A 5 s of simultaneously recorded signals are shown in Fig. 1.b. The SCG signals were then filtered using a low-pass filter with a cut-off of 100 Hz to remove the respiratory noise, which mostly has energy above this cut-off frequency [22]. Matlab (R2015b, The MathWorks, Inc, Natick, MA) was used to process all signals.

## C. SCG Segmentation and Grouping based on Respiration

The SCG events in each signal were found using a matched filtering with a template consisting of a previously identified SCG. The filtering algorithm was obtained from [23]. The matched filter coefficients, $w(t)$, were calculated as

$$w(t) = l(L - t + 1) \qquad (1)$$

where $l(t)$, $L$, and $t = 1,2,...,L$ were the library template (an SCG event manually chosen by the user), number of sample points in the template, and the coefficient index, respectively. The filter output, $y(t)$, was then calculated as

$$y(t) = w(t) * x(t) \qquad (2)$$

where $x(t)$ was the raw SCG signal. The filter output had maximums at locations that the raw SCG signal best matched the template. The envelope of the filter output was found using Hilbert transform. The peaks of this envelope signal with an amplitude above a certain threshold were then identified. The indices of the peaks were then used to determine the location of the SCG events. Identified SCG events were checked manually to confirm the absence of distorted SCG (for example due to motion artifacts). SCG events were then divided into two groups using two different respiratory criteria. First, the respiratory flow rate was used to group the SCG events into inspiratory and expiratory groups corresponding to positive and negative respiratory flows, respectively. Similarly, the lung volume was used to group SCG events. Here the average lung volume was first calculated, and low and high lung volumes (i.e., LLV and HLV) were defined as those below and above the mean lung volume. The SCG events were then labeled as LLV and HLV events, depending on if they occurred during LLV or HLV, respectively. The two grouping methods were then compared to determine which criterion is more effective in grouping similar SCG events. The details of quantifying the SCG event similarity and effectiveness of the grouping criteria are described in the next section.

TABLE I. OVERVIEW OF THE SUBJECTS' CHARACTERISTICS (MEAN ± SD).

| | |
|---|---|
| Age (years) | 24.3 ± 5.0 |
| Height (cm) | 170.8 ± 8.2 |
| Weight (kg) | 78.7 ± 13.0 |
| Heart rate (bpm) | 66.6 ± 9.0 |
| BMI (kg/m$^2$) | 26.9 ± 3.4 |
| Number of subjects | 7 |

*D. Grouping Criteria Effectiveness*

After separating the SCG events into two groups (e.g., inspiratory and expiratory), they were aligned in time (by minimizing the cross-correlation function), and an ensemble average SCG was calculated for each group separately. Fig. 2 shows the ensemble average of SCG events during LLV and HLV. To quantify the dissimilarity of each SCG with respect to the two groups, the difference between each SCG waveform and the average waveform of both groups was calculated. Then the RMS (root-mean-square) of these differences were determined (Eq. 3). This quantity was then divided by the RMS amplitude of the average waveform for each group (Eq. 4).

$$DRMS_{SCG_{i,j}} = RMS(SCG_{i,j} - SCG_{avg,j}) \quad (3)$$

$$RMS_{SCG_{i,j}} = \left| \frac{DRMS_{SCG_{i,j}}}{RMS_{SCG_{avg,j}}} \right| \times 100 \quad (4)$$

where $i \in [1, ..., number\ of\ events\ in\ each\ group]$, $j$ is the group (i.e., inspiration or expiration), and

$$RMS_{SCG_{avg,j}} = RMS(SCG_{avg,j}) \quad (5)$$

where $SCG_{avg,j}$ is the ensemble averaged SCG event of group $j$.

The average dissimilarity of grouped SCG events was calculated as,

$$\overline{RMS}_{SCG_j} = \sum_i RMS_{SCG_{i,j}} \quad (6)$$

This is calculated within the same group as well as with respect to the alternate group. For example, for events that were grouped as inspiratory, their average dissimilarity was calculated with respect to inspiratory (i.e., same group) and expiratory (i.e., alternative group), separately. The difference between these two average dissimilarities is indicative of how well was the grouping and can be calculated from,

$$RD_{FR_i} = \left(\overline{RMS}_{SCG_{i\in j,insp}} - \overline{RMS}_{SCG_{i\in j,exp}}\right) / \overline{RMS}_{SCG_{i\in j,insp}} \quad (7.a)$$

where $RD_{FR}$ is the normalized difference of mean dissimilarity of inspiratory events with respect to inspiratory and expiratory groups, respectively. The same dissimilarity difference was calculated for expiratory events.

Another grouping choice for SCG events that was tested in the current study was based on low and high lung volume, LLV and HLV, respectively. Here, dissimilarities were also calculated to determine the dissimilarity of each SCG event group with respect to its own group and the alternative group. For example, for LLV SCG events, the difference between average dissimilarities relative to LLV and HLV groups was calculated from,

$$RD_{LV_i} = \left(\overline{RMS}_{SCG_{i\in j,LLV}} - \overline{RMS}_{SCG_{i\in j,HLV}}\right) / \overline{RMS}_{SCG_{i\in j,LLV}} \quad (7.b)$$

To determine which grouping criterion (i.e., inspiration vs expiration or LLV vs HLV) provide better grouping of SCG events, the difference in the average dissimilarity was compared. For example, the $RD_{FR}$ and $RD_{LV}$ were compared for each subject. This will help determine whether the respiratory flow rate or lung volume more effectively separate SCG events.

III. RESULTS

The mean dissimilarity of inspiratory events with respect to same or alternative group (i.e., inspiratory and expiratory groups, respectively) are listed in Table II (Column 2 and 3, respectively). Column 4 shows the number of events in the group. The same information is listed for the expiratory group in columns 5, 6, and 7, respectively. The difference in dissimilarity between alternate and same group is listed in columns 8 and 9, respectively, where positive values indicate more dissimilarity with the alternate group compared to the same group. The difference was positive in 6 out of 7 subjects. Hence it can be concluded that in most subjects, the mean dissimilarity within the same group was smaller than that for the alternative group, indicating proper grouping. The fact that two different morphologies of SCG can be separated based on

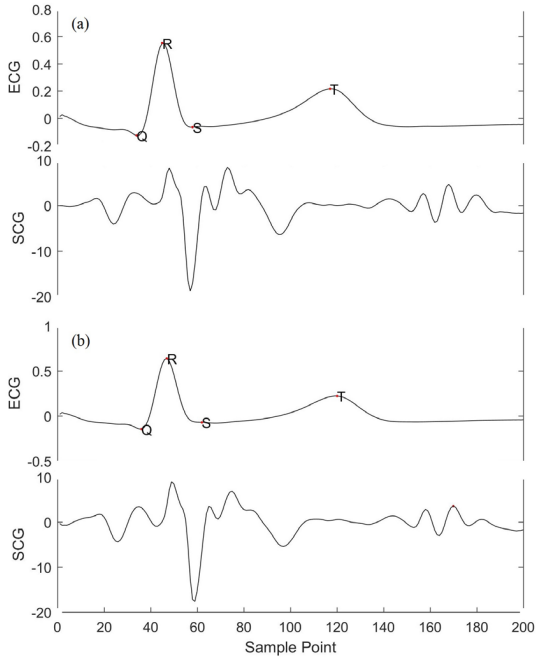

Fig. 2. (a) Ensemble averaged ECG in the top panel and ensemble averaged SCG in the bottom panel during the high lung volume, (b) Ensemble averaged ECG in the top panel and ensemble averaged SCG in the bottom panel during the low lung volume.

TABLE II.     RMS BETWEEN SCG EVENTS IN THE INSPIRATORY AND EXPIRATORY GROUPS AND THE ENSEMBLE AVERAGED INSPIRATORY/EXPIRATORY SCG EVENT (EQUATION 6). THE VALUES ARE SHOWN AS MEAN ± SD. THE NUMBER OF SCG EVENTS IN EACH GROUP IS SHOWN IN PARENTHESIS. THE LAST COLUMN SHOWS THE RELATIVE DIFFERENCES IN PERCENTAGE.

| Subject # | RMS between inspiratory events and … | | | RMS between expiratory events and … | | | Relative Difference (%) | |
|---|---|---|---|---|---|---|---|---|
| | Averaged inspiratory SCG $\overline{RMS}_{SCG_{insp,insp}} \pm SD$ | Averaged expiratory SCG $\overline{RMS}_{SCG_{insp,exp}} \pm SD$ | | Averaged inspiratory SCG $\overline{RMS}_{SCG_{exp,insp}} \pm SD$ | Averaged expiratory SCG $\overline{RMS}_{SCG_{exp,exp}} \pm SD$ | | $RD_{FR_{insp}}$ | $RD_{FR_{exp}}$ |
| 1 | 25.0252 ± 6.0923 | 33.2976 ± 7.3152 | (53) | 28.4763 ± 7.5722 | 24.0721 ± 7.2852 | (55) | 33.06 | 18.29 |
| 2 | 42.1056 ± 12.3828 | 46.6644 ± 12.0475 | (38) | 50.7218 ± 11.9698 | 47.8863 ± 13.2326 | (54) | 10.83 | 5.92 |
| 3 | 47.7511 ± 13.4712 | 61.4332 ± 15.1620 | (37) | 56.7565 ± 20.0213 | 49.7977 ± 21.2623 | (35) | 28.65 | 13.97 |
| 4 | 45.8798 ± 13.6270 | 64.3130 ± 18.7429 | (28) | 65.5481 ± 14.5013 | 43.6113 ± 15.7970 | (55) | 40.18 | 50.30 |
| 5 | 33.5099 ± 9.16038 | 36.2629 ± 7.9248 | (46) | 36.3790 ± 10.1633 | 31.8645 ± 8.5258 | (28) | 8.21 | 14.17 |
| 6 | 32.6761 ± 2.1323 | 45.2334 ± 8.3686 | (26) | 48.2500 ± 17.0370 | 48.9503 ± 11.3665 | (53) | 38.43 | -1.43 |
| 7 | 33.2406 ± 8.8591 | 44.2626 ± 7.9633 | (36) | 40.0089 ± 8.6755 | 33.1932 ± 8.1408 | (43) | 33.16 | 20.53 |

TABLE III.     RMS BETWEEN SCG EVENTS IN THE HLV AND LLV GROUPS AND THE ENSEMBLE AVERAGED HLV/LLV SCG EVENT (EQUATION 6). THE VALUES ARE SHOWN AS MEAN ± SD. THE NUMBER OF SCG EVENTS IN EACH GROUP IS SHOWN IN PARENTHESIS. THE LAST COLUMN SHOWS THE RELATIVE DIFFERENCES IN PERCENTAGE.

| Subject # | RMS between LLV events and … | | | RMS between HLV events and … | | | Relative Difference (%) | | |
|---|---|---|---|---|---|---|---|---|---|
| | Averaged LLV SCG $\overline{RMS}_{SCG_{LLV,LLV}} \pm SD$ | Averaged HLV SCG $\overline{RMS}_{SCG_{LLV,HLV}} \pm SD$ | n | Averaged LLV SCG $\overline{RMS}_{SCG_{HLV,LLV}} \pm SD$ | Averaged HLV SCG $\overline{RMS}_{SCG_{HLV,HLV}} \pm SD$ | n | $RD_{LV_{LLV}}$ | $RD_{LV_{HLV}}$ | |
| 1 | 22.4070 ± 5.6409 | 34.1765 ± 9.4193 | (53) | 31.4550 ± 5.4819 | 24.9368 ± 6.8360 | (58) | 52.52 | 26.14 | + |
| 2 | 46.1731 ± 11.3189 | 49.5236 ± 14.7105 | (46) | 63.2857 ± 11.2033 | 43.2033 ± 13.2348 | (42) | 7.26 | 46.48 | + |
| 3 | 47.5796 ± 12.7507 | 87.6964 ± 13.8409 | (32) | 86.4990 ± 18.5154 | 51.9888 ± 20.0642 | (39) | 84.31 | 66.38 | + |
| 4 | 44.0150 ± 11.3531 | 77.7045 ± 21.2454 | (44) | 69.7121 ± 9.4823 | 34.0905 ± 12.6197 | (37) | 76.54 | 104.49 | + |
| 5 | 30.4948 ± 8.1303 | 44.8703 ± 7.6175 | (44) | 59.2533 ± 10.8211 | 25.3256 ± 7.1247 | (31) | 47.14 | 133.97 | + |
| 6 | 33.1170 ± 10.1451 | 63.0525 ± 13.6690 | (51) | 69.8478 ± 18.9298 | 37.1869 ± 10.4012 | (31) | 90.39 | 87.83 | + |
| 7 | 26.0518 ± 6.7255 | 60.4353 ± 13.9944 | (40) | 43.9254 ± 8.3870 | 34.5265 ± 7.9952 | (36) | 131.98 | 27.22 | + |

The "+" sign in the right column indicates that the dissimilarity results improved when lung volume was used instead of respiratory flow rate.

respiration is consistent with reports that the SCG morphology changes with different phases of respiration [17].

For one subject (subject #6), the expiratory events were more dissimilar to their group than the inspiratory group. For this subject, the two mean dissimilarities were close (~ 48.5).

Results of grouping SCG based on lung volume (i.e. LLV vs HLV) are presented in Table III, where the format is parallel to Table II. The SCG events in each group (LLV and HLV) were more similar to their own group.

The $RD_{FR}$ and $RD_{LV}$ values (listed in the last column of TABLE II and TABLE III, respectively) are a measure of how well SCG grouping was. These results show that the lung volume signal was more successful than the flow rate signal in grouping the SCG events into two different groups (Fig. 3) where the events in each group are morphologically similar to each other.

### IV. DISCUSSIONS

#### A. Intrathoracic Pressure and Heart Displacement

The chest wall expansion and downward movement of the diaphragm during inspiration causes a more negative intrathoracic pressure and a downward movement of the heart. The negative pressure increases the expansion of the right atrium, right ventricle and thoracic superior and inferior vena cava, which causes the intravascular and intra-cardiac pressures to fall. As a result, the transmural pressure (the difference between pressure inside the heart chamber and the intrathoracic pressure) increases. This causes a rise in cardiac chamber expansion, preload and stroke volume through the Frank-Starling mechanism. The opposite phenomena happens during expiration [21]. It can then be concluded that intrathoracic pressure variations due to respiration changes the heart chamber pressures, preload, stroke volume and stroke work [24]. These mechanical changes are expected to affect the heart muscle contractile movements and blood flow momentum which can manifest themselves as variations in SCG signal morphologies.

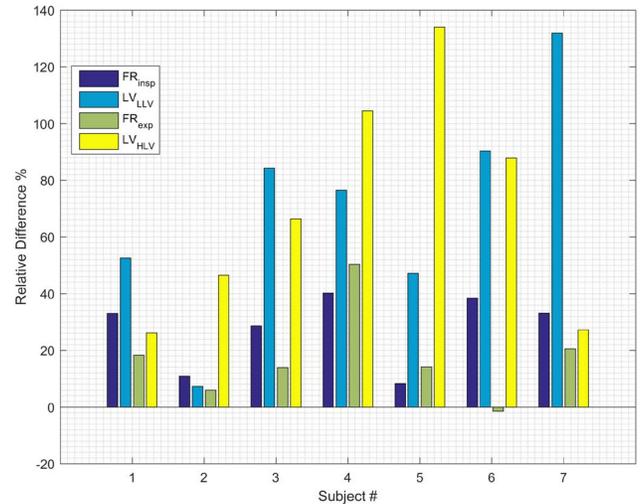

Fig. 3. Relative differences calculated from Eq. 7.a and 7.b for inspiratory, expiratory, LLV, and HLV SCG events.

*B. Limitations*

The primary limitation of this study was the small number of subjects that participated. Future studies need to enroll larger number of subjects from a diverse population including different age, gender, weight, race, and clinical status.

## V. CONCLUSIONS

The results of this study showed that the SCG demonstrated morphological differences during respiration. SCG events were grouped according to their waveform morphology. Two grouping criteria were implemented. One grouping relied on inspiratory vs. expiratory flow while the other relied on LLV vs HLV (which corresponds to high and low intrathoracic pressure). The second criterion resulted in more similarity within the SCG groups, suggesting that intrathoracic pressure variations can lead to detectable SCG morphology changes. Studying the effect of respiration allows separating SCG into groups with similar events. This reduces SCG waveform variability and enables more precise estimation of SCG characteristics. In addition, because respiration triggers known changes in physiological parameters (such as intrathoracic pressure, stroke volume, etc.), it allows studying the effects of these parameters on SCG. Such investigations can help enhance our understanding of SCG genesis, and explain SCG changes with cardiac pathology. Future studies may perform comparisons between the spectral characteristics of two groups of SCG (e.g., LLV vs HLV) as this might reveal further useful SCG characteristics and may contribute to further elucidate SCG genesis. In addition, artificial intelligence methods such as neural networks or support vector machines might be used to classify the SCG events into two groups. An ongoing study [25] aims at developing classification algorithms for this purpose.

ACKNOWLEDGMENTS

This study was supported by NIH R44HL099053.